Doping control of magnetic anisotropy for stable antiskyrmion formation in schreibersite (Fe,Ni)$_3$P with $S_4$ symmetry

*Kosuke Karube\*, Licong Peng, Jan Masell, Mamoun Hemmida, Hans-Albrecht Krug von Nidda, István Kézsmárki, Xiuzhen Yu, Yoshinori Tokura, Yasujiro Taguchi*

K. Karube, L. C. Peng, X. Z. Yu, Y. Taguchi
RIKEN Center for Emergent Matter Science (CEMS), Wako 351-0198, Japan.
E-mail: kosuke.karube@riken.jp

J. Masell
RIKEN Center for Emergent Matter Science (CEMS), Wako 351-0198, Japan.
Institute of Theoretical Solid State Physics, Karlsruhe Institute of Technology (KIT), 76049 Karlsruhe, Germany.

M. Hemmida, H.-A. Krug von Nidda, I. Kézsmárki
Experimental Physics V, University of Augsburg, 86135 Augsburg, Germany.

Y. Tokura,
RIKEN Center for Emergent Matter Science (CEMS), Wako 351-0198, Japan.
Department of Applied Physics, University of Tokyo, Bunkyo-ku 113-8656, Japan.
Tokyo College, University of Tokyo, Bunkyo-ku 113-8656, Japan.

Magnetic skyrmions, vortex-like topological spin textures, have attracted much interest in a wide range of research fields from fundamental physics to spintronics applications. Recently, growing attention has also been paid to antiskyrmions emerging in opposite topological charge in non-centrosymmetric magnets with $D_{2d}$ or $S_4$ symmetry. In these magnets, complex interplay among anisotropic Dzyaloshinskii-Moriya interaction, uniaxial magnetic anisotropy, and magnetic dipolar interactions generates a variety of magnetic structures. However, the relation between the stability of antiskyrmions and these magnetic interactions remains to be elucidated. In this work, we control the uniaxial magnetic anisotropy of schreibersite (Fe,Ni)$_3$P with $S_4$ symmetry by doping and investigate its impact on the stability of antiskyrmions. Our magnetometry study, supported by ferromagnetic resonance spectroscopy, shows that the variation of the Ni content and slight doping with 4$d$ transition metals considerably change the magnetic anisotropy. In particular, doping with Pd induces easy-axis anisotropy, giving rise to formation of antiskyrmions, while a temperature-induced spin reorientation is observed in a Rh-doped compound. In combination with Lorentz transmission electron microscopy and micromagnetic simulations, we quantitatively analyze the stability of antiskyrmion as functions of uniaxial anisotropy and demagnetization energy, and demonstrate that subtle balance between them is necessary to stabilize the antiskyrmions.

1. Introduction

Vortex-like spin swirling objects, termed magnetic skyrmions and characterized by an integer topological winding number $N_{sk}$, have been extensively studied in the last decade, both in the fields of fundamental science and applications to spintronics devices.[1,2] One of the well-known mechanisms of skyrmion formation is the competition between ferromagnetic exchange interaction and the Dzyaloshinskii-Moriya interaction (DMI) arising from the lack of inversion symmetry. Skyrmions with helical (Bloch type) and cycloidal (Néel type) spin configurations have been observed in non-centrosymmetric magnets with chiral ($T$ or $O$ class) and polar ($C_{nv}$ class) structures, respectively.[3–9] Recently, a new topological spin texture, the antiskyrmion, with opposite sign of $N_{sk}$ for the same polarity has attracted much attention. The antiskyrmion consists of both Bloch and Néel walls with the opposite helicities along two orthogonal axes, and its formation is attributed to the anisotropic DMI present in non-centrosymmetric tetragonal crystals belonging to the $D_{2d}$ or $S_4$ symmetry group both containing four-fold rotoinversion ($\bar{4}$).[10,11] In real materials, antiskyrmions were first found in Heusler compounds with $D_{2d}$ symmetry, $Mn_{1.4}PtSn$ and $Mn_{1.4}Pt_{0.9}Pd_{0.1}Sn$.[12,13] More recently, the formation of antiskyrmions was also found in $Fe_{1.9}Ni_{0.9}Pd_{0.2}P$ [Pd-doped (Fe,Ni)$_3$P] with $S_4$ symmetry[14] and Fe/Gd-based multilayers[15]. Lorentz transmission electron microscopy (LTEM) observation for thin plates of these $D_{2d}$ and $S_4$ magnets shows that the antiskyrmions are square shaped, and transform into bullet-shaped non-topological bubbles and elliptically deformed Bloch skyrmions, depending on magnetic fields, temperature, and lamella thickness.[13,14,16]

According to numerical simulations, magnetic dipolar interaction (demagnetization energy) plays a dominant role in the formation of square-shaped antiskyrmions;[13,16,17] Néel walls (Bloch lines) at the corners of an antiskyrmion tend to shrink so as to reduce the magnetic volume charge.[18] Magnetic force microscopy studies have shown that the size of antiskyrmions increases significantly as the crystal thickness is increased,[14,19] and anisotropic fractal-like domains with $\bar{4}$ symmetry are induced near the surface of bulk crystals[14,20] by the interplay between DMI, uniaxial magnetic anisotropy and dipolar interactions to reduce the magnetic charge on the surface.[21]

Despite these findings in the two families of antiskyrmion-hosting magnets, the relation between the stability of antiskyrmions and magnetic interactions (DMI, magnetic anisotropy, and dipolar interaction) is still unclear due to the lack of experimental studies where the relative strength of magnetic interactions is tuned systematically. In this work, we succeeded in controlling the magnetic anisotropy of schreibersite (Fe,Ni)$_3$P with $S_4$ symmetry by chemical doping to observe the formation of stable antiskyrmions. First, we searched for the appropriate solid solution of Fe and Ni to reduce the easy-plane anisotropy. Next, using 4$d$ transition metals with strong spin-orbit coupling, we further modified the magnetic anisotropy to easy-axis type, leading to the formation of stable antiskyrmions and skyrmions. We finally identified the stable region of antiskyrmions and skyrmions on the plane of uniaxial magnetic anisotropy and demagnetization energy, thereby demonstrating the subtle balance between them to give rise to the antiskyrmion spin texture.

## 2. Results and Discussions

### 2-1. Structual properties

We synthesized bulk single crystals of schreibersite (Fe$_{1-x}$Ni$_x$)$_3$P ($0 \leq x \leq 0.66$) and those doped with small amounts of 4$d$ transition metals (Ru, Rh and Pd) listed in **Table 1** by a self-flux method (see Note S1 and Table S1 for details of sample preparation). Using powder X-ray diffraction, we confirmed that all the compounds crystalize in a non-centrosymmetric tetragonal structure with the space group of $I\bar{4}$ (No. 82, $S_4^2$) as shown in **Figure 1**a (see Note S2, Figure S1, and Figure S2 for details). The tetragonal lattice constants ($a$, $c$) obtained by Rietveld analysis are plotted in Figure 1b. The parameter $a$ decreases linearly with increasing Ni concentration, while $c$ is almost unchanged below $x = 0.47$, in good agreement with a previous study.[22] When a small amount of Pd is substituted for Ni, both $a$ and $c$ increase linearly following Vegard's law. Similar increase in the lattice constants is also observed with Ru and Rh doping.

**Table 1.** Lattice constants ($a$, $c$) at room temperature, magnetic transition temperature ($T_c$), saturation magnetization ($M_s$) and uniaxial anisotropy constant ($K_u$) at 300 K and 2 K for (Fe,Ni)$_3$P and doped with 4$d$ transition metals.

| Composition | $a$ [Å] | $c$ [Å] | $T_c$ [K] | $M_s$ (300 K) [kA/m] | $M_s$ (2 K) [kA/m] | $K_u$ (300 K) [kJ/m$^3$] | $K_u$ (2 K) [kJ/m$^3$] |
|---|---|---|---|---|---|---|---|
| Fe$_3$P | 9.1018(8) | 4.4587(4) | 681 | 1080 | 1160 | −647 | −848 |
| (Fe$_{0.82}$Ni$_{0.18}$)$_3$P | 9.0713(4) | 4.4627(2) | 557 | 801 | 925 | −98.4 | −205 |
| (Fe$_{0.63}$Ni$_{0.37}$)$_3$P | 9.0425(5) | 4.4607(3) | 412 | 518 | 743 | −6.73 | −38.8 |
| (Fe$_{0.53}$Ni$_{0.47}$)$_3$P | 9.0266(5) | 4.4561(3) | 334 | 306 | 632 | −4.06 | −44.6 |
| (Fe$_{0.34}$Ni$_{0.66}$)$_3$P | 9.0006(5) | 4.4429(3) | 152 | - | 381 | - | −62.4 |
| (Fe$_{0.59}$Ni$_{0.32}$Ru$_{0.09}$)$_3$P | 9.0934(6) | 4.4907(3) | 337 | 363 | 693 | −5.65 | −53.2 |
| (Fe$_{0.60}$Ni$_{0.32}$Rh$_{0.08}$)$_3$P | 9.0931(7) | 4.4855(4) | 394 | 503 | 731 | 1.89 | −20.1 |
| (Fe$_{0.63}$Ni$_{0.33}$Pd$_{0.04}$)$_3$P | 9.0797(6) | 4.4745(3) | 399 | 511 | 750 | 22.2 | 51.3 |
| (Fe$_{0.63}$Ni$_{0.30}$Pd$_{0.07}$)$_3$P | 9.1122(7) | 4.4865(4) | 398 | 474 | 707 | 27.9 | 83.8 |
| (Fe$_{0.62}$Ni$_{0.29}$Pd$_{0.09}$)$_3$P | 9.1199(6) | 4.4895(3) | 392 | 451 | 704 | 33.5 | 114 |

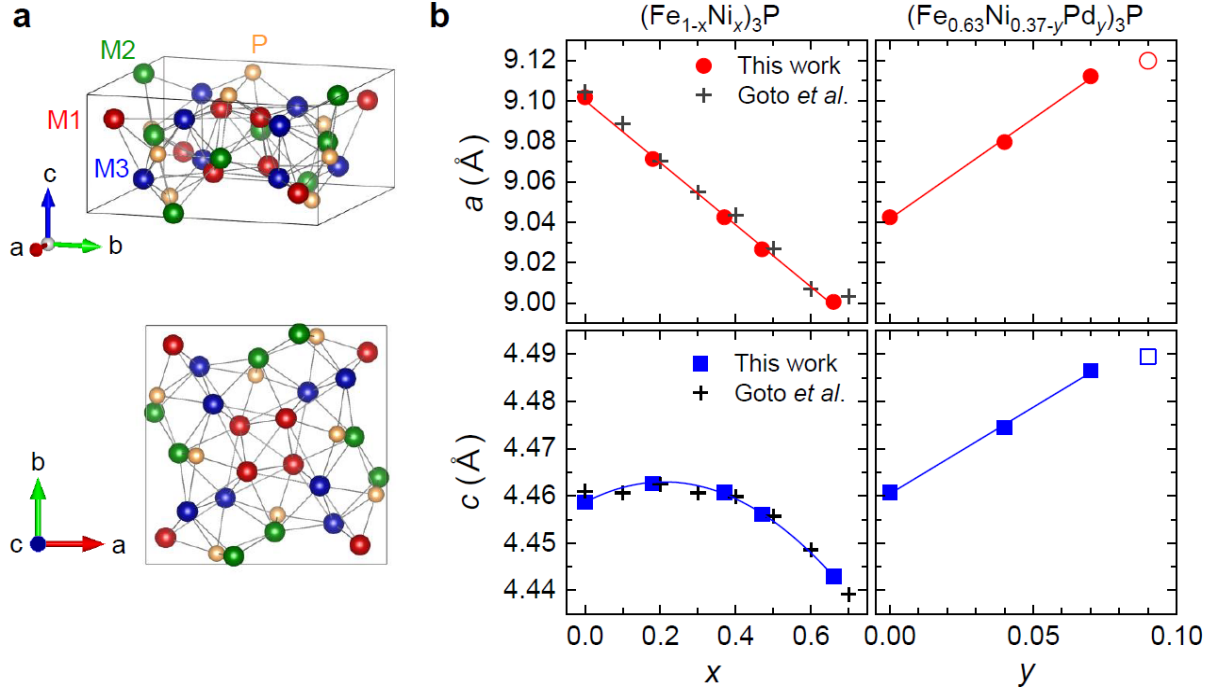

**Figure 1.** Structural properties of $(Fe_{1-x}Ni_x)_3P$. a) Crystal structure of $M_3P$ (M: transition metal) with the non-centrosymmetric tetragonal space group of $I\bar{4}$. The structure viewed from the $c$-axis is shown at the bottom. Three crystallographically inequivalent M sites are denoted as M1 (red), M2 (green) and M3 (blue). b) Lattice constants $a$ and $c$ at room temperature obtained from powder X-ray diffraction and Rietveld analysis are plotted as a function of $x$ of $(Fe_{1-x}Ni_x)_3P$ and $y$ of $(Fe_{0.63}Ni_{0.37-y}Pd_y)_3P$. The data for $(Fe_{1-x}Ni_x)_3P$ from literature are also plotted with cross symbols.[22] The data points of $(Fe_{0.62}Ni_{0.29}Pd_{0.09})_3P$ plotted with open symbols deviate from the linear fit due to the slightly lower Fe concentration.

## 2-2. Magnetometry

To characterize bulk magnetic properties, magnetization ($M$) was measured under magnetic fields applied parallel to the [110] and [001] axes, being perpendicular and parallel to the $\bar{4}$ axis, respectively. For the magnetization measurements, single crystals were cut into a rectangular shape (Figure S3) so that the demagnetization factors in the [110] and [001] directions were the same. By adopting this sample shape with the same demagnetization factors, the uniaxial magnetic anisotropy constant ($K_u$) is directly obtained, without explicitly taking the demagnetization effect into account, from the difference between the Helmholtz magnetic free energy along the [110] and [001] axes,[23]

$$K_u = \int_0^{M_s} \left[ H_{[110]}(M) - H_{[001]}(M) \right] dM \qquad (1)$$

where $M_s$ is the saturation magnetization, and $H_i(M)$ represents the magnetic field along the $i$-axis. Therefore, $K_u$ is equal to the area enclosed by the magnetization curves along the two directions. In the present case, the magnetization process along the easy axis is dominated by the displacement of domain walls, but the contribution of this process is excluded in Equation (1). The sign of $K_u$ indicates the direction of the anisotropy; $K_u > 0$ for the easy-axis anisotropy while $K_u < 0$ for the easy-plane type.

The temperature dependence $M(T)$ at 0.01 T and the magnetic field variation $M(H)$ at 2 K in $(Fe_{1-x}Ni_x)_3P$ are presented in **Figure 2**. $Fe_3P$ exhibits a ferromagnetic transition around 680 K and shows strong easy-plane anisotropy, as indicated by smaller $M$ values and larger saturation field in the [001] axis than those in the [110] axis (Figure 2a,f). Partial substitution of Fe with Ni lowers the magnetic transition temperature $T_c$ (Figure 2b-e) and decreases the saturation value of magnetization $M_s$ (Figure 2g-j). Furthermore, the difference between the magnetization curves along the [110] and [001] axes becomes smaller as the Ni substitution proceeds, and is the smallest at the Ni concentration of 37%. $M(T)$ of $(Fe_{0.34}Ni_{0.66})_3P$ decreases below 40 K (Figure 2e) probably due to the spin glass behavior of the diluted Fe moments.

The magnetic parameters, $T_c$, $M_s$, and $K_u$, obtained from the magnetization measurements using Equation (1) are plotted in Figure 2k-m as a function of Ni concentration $x$. The monotonous decrease in $T_c$ and $M_s$ with increasing $x$ is in accord with the previous report on polycrystalline samples.[22] While there have been several studies on the magnetic properties of $(Fe_{1-x}M_x)_3P$ (M: Cr, Mn, Co, Ni),[22,24–26] magnetocrystalline anisotropy has been reported only for $(Fe_{1-x}Co_x)_3P$.[27] As shown in Figure 2m, the large negative value of $K_u$ for $Fe_3P$ (−848 kJ/m$^3$ at 2 K) determined in the present study coincides with the previous report.[27] The absolute value of $K_u$ is rapidly suppressed by Ni substitution, reaching a minimum at $x = 0.37$ (−39 kJ/m$^3$ at 2 K). Nevertheless, the easy-plane anisotropy ($K_u < 0$) persists throughout the whole Ni concentration range in the present study of $(Fe_{1-x}Ni_x)_3P$.

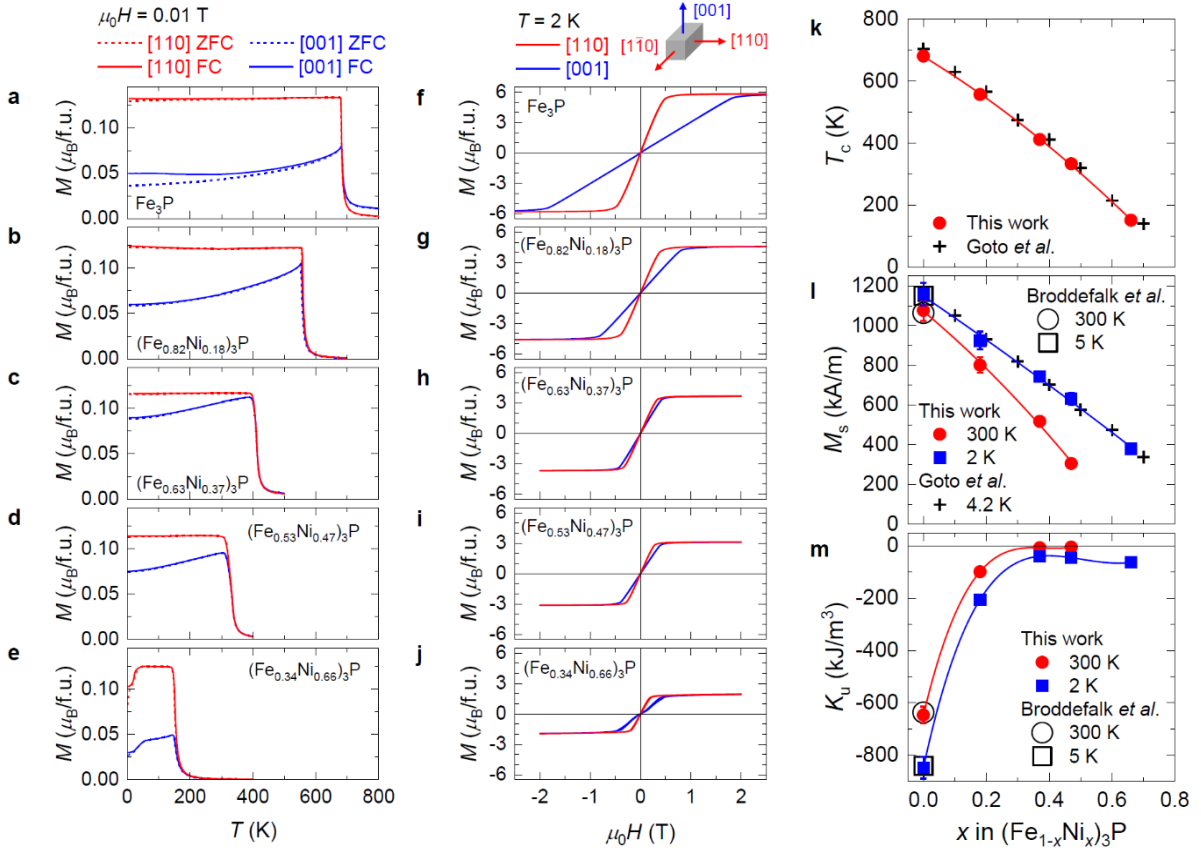

**Figure 2.** Magnetic properties of $(Fe_{1-x}Ni_x)_3P$. a-e) Temperature ($T$) dependence of magnetization $M$ under the magnetic field $\mu_0 H = 0.01$ T for (a) $Fe_3P$, (b) $(Fe_{0.82}Ni_{0.18})_3P$, (c) $(Fe_{0.63}Ni_{0.37})_3P$, (d) $(Fe_{0.53}Ni_{0.47})_3P$, and (e) $(Fe_{0.34}Ni_{0.66})_3P$. The data of field cooling (FC) and field warming after a zero-field cooling (ZFC) are denoted with solid and dotted lines, respectively. f-j) Magnetic field dependence of magnetization at 2 K for the same compositions as shown in the panels a-e. Magnetizations for the [110] and [001] axes are shown by red and blue lines, respectively. The schematic figure of a rectangular bulk single crystal used for the measurements is shown at the top of panel f. k-m) Ni concentration $x$ dependence of (k) ferromagnetic transition temperature $T_c$, (l) saturation magnetization $M_s$, and (m) uniaxial anisotropy constant $K_u$ at 300 K and 2 K obtained from the magnetization measurements. The value of $M_s$ was determined as the magnetization at $\mu_0 H = 3$ T for $Fe_3P$, 2 T for $x = 0.18$, and 1 T for $x \geq 0.37$. For the conversion of $M_s$ and $K_u$ to unit volume, the lattice constants obtained by powder X-ray diffraction at room temperature were used. $T_c$ and $M_s$ of polycrystalline $(Fe_{1-x}Ni_x)_3P$ in the literature are indicated by cross symbols.[22] $M_s$ and $K_u$ of $Fe_3P$ at 300 K and 5 K in the literature are shown with open symbols.[27]

Having thus established that the strong in-plane magnetic anisotropy (negative $K_u$) in $(Fe_{1-x}Ni_x)_3P$ can be appreciably reduced around $x \sim 0.4$ while keeping high $T_c$ above room temperature, next we aim to further control $K_u$ and to reverse its sign via doping with $4d$ transition metals. **Figure 3** shows the effect of $4d$ transition-metal doping on the magnetic properties. In the Ru-doped compound, $(Fe_{0.59}Ni_{0.32}Ru_{0.09})_3P$ (Figure 3a,f), weak easy-plane anisotropy is observed in the whole temperature range below $T_c \sim 340$ K, as in the case of (Fe,Ni)$_3$P. On the other hand, the Rh-doped sample, (Fe$_{0.60}$Ni$_{0.32}$Rh$_{0.08}$)$_3$P (Figure 3b,g), shows a complex temperature dependence. Below $T_c \sim 400$ K, the magnetization along the [110] axis increases gradually upon cooling, whereas that along the [001] axis is almost independent of temperature. The magnetization along the [110] axis saturates at 100 K, below which the magnetization along the [001] axis slightly decreases. This behavior is a typical indicator of spin reorientation from the $c$-axis to the $ab$-plane, which is often seen in uniaxial ferromagnets such as MnBi,[28] Fe$_3$Sn$_2$,[29,30] and Mn$_{1.4}$PtSn.[31] The spin reorientation temperature in the Rh-doped compound is determined to be $T_{SR} \sim 100$ K where the kink-like anomaly is observed. As for another dopant Pd, 4% doped (Fe$_{0.63}$Ni$_{0.33}$Pd$_{0.04}$)$_3$P (Figure 3c,h) does not show spin reorientation and exhibits weak easy-axis anisotropy in the whole temperature range below $T_c \sim 400$ K. In Pd 7% doped (Fe$_{0.63}$Ni$_{0.30}$Pd$_{0.07}$)$_3$P (Figure 3d,i) and 9% doped (Fe$_{0.62}$Ni$_{0.29}$Pd$_{0.09}$)$_3$P (Figure 3e,j), the easy-axis anisotropy is more clearly seen than in the Pd 4% doped sample.

The values of $M_s$ and $K_u$ in the investigated compounds are plotted against temperature in Figure 3k and 3l. While the $M_s$ values are similar for the different compositions, the sign and the absolute value of $K_u$ considerably change with the 4$d$ metal dopant species. In particular, easy-axis anisotropy ($K_u > 0$) is induced by small amount of Pd doping, and the value of $K_u$ increases with decreasing temperature and increasing the Pd concentration. The temperature dependence of $K_u$ and $M_s$ in the Pd 7% and 9% doped compounds are found to obey the relation $K_u(T) \propto [M_s(T)]^{2.7}$ (Figure S4). The power law with an exponent close to 3 agrees with the theoretical models for uniaxial anisotropy.[32-34] For the Pd 7% compound that shows $M_s \sim 474$ kA/m and $K_u \sim 28$ kJ/m$^3$ at room temperature, the quality factor is estimated to be $Q \sim 0.2$, where $Q$ is defined as $Q = K_u/(\mu_0 M_s^2/2)$, the ratio of the uniaxial anisotropy constant to the maximum of demagnetization energy. These $K_u$ and $Q$ values are small as compared with those of the industrially-important commercial permanent magnets, such as (Sr,Ba)Fe$_{12}$O$_{19}$, Nd$_2$Fe$_{14}$B, SmCo$_5$,[35] and the recently reported Heusler compound Mn$_{1.4}$PtSn ($K_u \sim 171$ kJ/m$^3$, $Q \sim 1.7$ at room temperature).[36] The values of $K_u$ for the Rh-doped compound are in-between those for the Ru- and Pd-doped ones, and change from very small positive values at room temperature to negative values at low temperatures, leading to the temperature-induced spin reorientation. The significant change in the magnetic anisotropy from easy-plane to easy-axis with the Pd doping indicates the importance of enhanced spin-orbit coupling in the 4$d$ element. The systematic variation in the magnetic anisotropy with change in the dopant Ru, Rh, Pd is probably dominated by band filling of 4$d$ orbitals, or position of the Fermi level.

While the systematic change of $K_u$ with composition and temperature was identified by the magnetometry study using Equation (1), we performed ferromagnetic resonance (FMR) spectroscopy experiments on the Pd 7% compound to determine $K_u$ by another method, and thus further validate our magnetometry-based approach for the quantification of anisotropy.

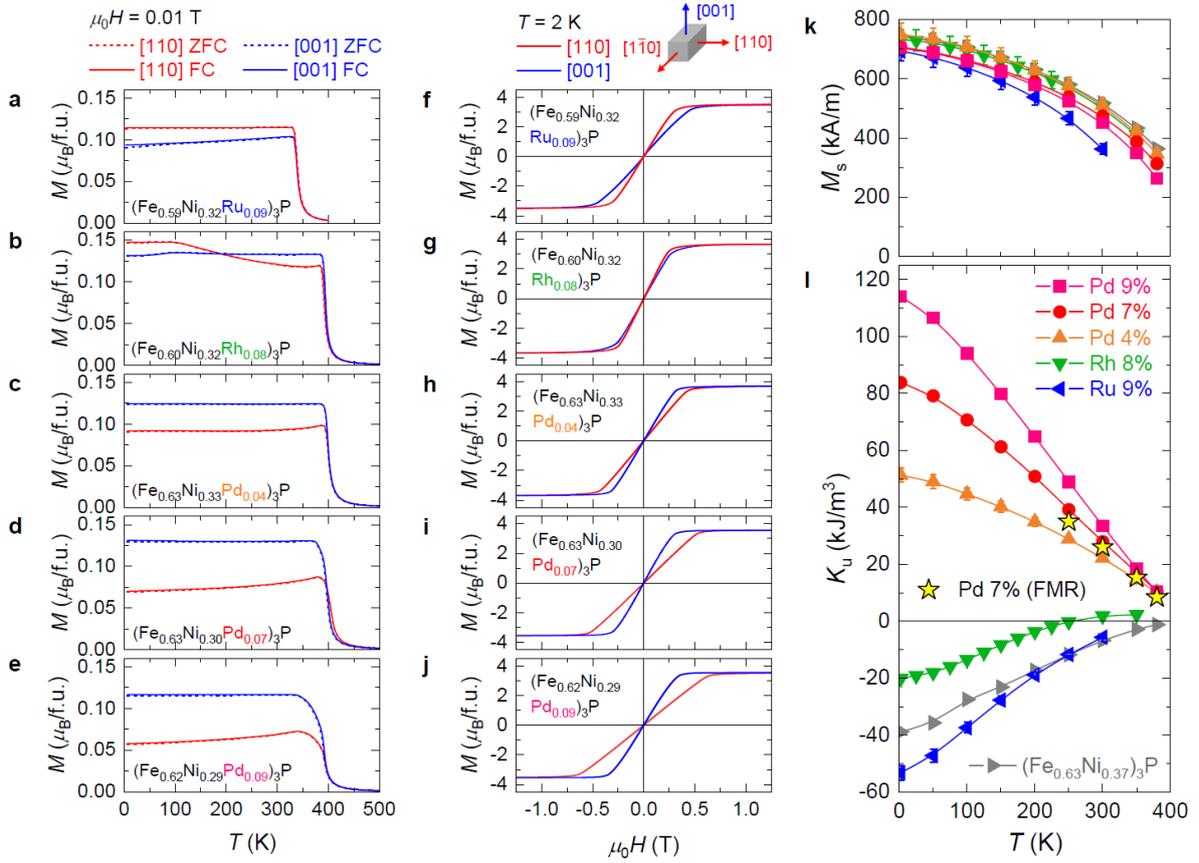

**Figure 3.** Magnetic properties of 4$d$-metal doped compounds. a-e) Temperature dependence of magnetization under the magnetic field of $\mu_0 H = 0.01$ T for (a) $(Fe_{0.59}Ni_{0.32}Ru_{0.09})_3P$, (b) $(Fe_{0.60}Ni_{0.32}Rh_{0.08})_3P$, (c) $(Fe_{0.63}Ni_{0.33}Pd_{0.04})_3P$, (d) $(Fe_{0.63}Ni_{0.30}Pd_{0.07})_3P$ (equivalent to the notation $Fe_{1.9}Ni_{0.9}Pd_{0.2}P$ as adopted in our earlier work[14]), and (e) $(Fe_{0.62}Ni_{0.29}Pd_{0.09})_3P$. f-j) Magnetic field dependence of magnetization at 2 K for the same compositions as shown in the panels a-e. The magnetization data for Pd 9% are reproduced with permission from our earlier work.[14] (Copyright 2021, Springer Nature). The schematic figure of a rectangular bulk single crystal used for the measurements is shown at the top of panel f. k, l) Temperature dependence of (k) $M_s$ and (l) $K_u$ obtained from the magnetization measurements for the various compositions. The value of $M_s$ was determined as the magnetization at 1 T for all the compounds. The values of $K_u$ in $(Fe_{0.63}Ni_{0.30}Pd_{0.07})_3P$ obtained from the FMR measurement are also plotted with yellow star symbols.

## 2-3. Ferromagnetic resonance spectroscopy

We carried out FMR spectroscopy measurements at 9.4 GHz on a cylindrical disk prepared from a Pd 7% crystal and shown in the inset of **Figure 4**b. Figure 4a displays representative field-swept FMR spectra recorded at room temperature for various orientations of the magnetic field applied in the plane of the disk, which contains both the [110] and [001] axes. The maxima of the microwave absorption curves correspond to the FMR field, $H_{res}$, whose angular dependence is displayed in Figure 4b for a full rotation of the field in the plane of the disk. The minimum of $H_{res}$ is observed for magnetic field along the [001] axis, while the maximum is reached for field parallel to the [110] axis, clearly demonstrating the easy-axis nature of the anisotropy with the [001] direction being the easy axis. Due to the cylindrical shape of the sample, the demagnetization factor is unchanged upon the field rotation, hence the angular dependence of $H_{res}$ is solely governed by $K_u$. Similar $H_{res}(\theta)$ curves were recorded at 250 K, 350 K and 380 K. The $K_u$ values obtained by fitting these curves, as described in the Methods section and in Ref.[37–39], are plotted in Figure 3l. These values are in excellent agreement with those obtained from the analysis of the magnetization, supporting the fully quantitative determination of $K_u$ in the present work.

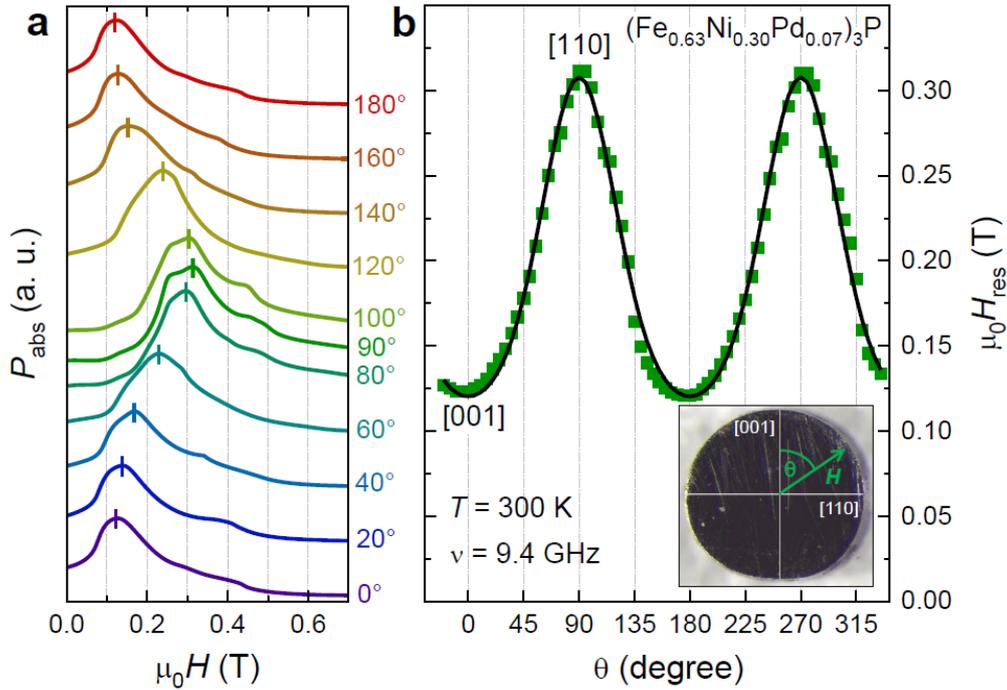

**Figure 4.** Ferromagnetic resonance (FMR) spectroscopy on $(Fe_{0.63}Ni_{0.30}Pd_{0.07})_3P$ at 300 K. a) FMR spectra, representing the absorbed power $P_{abs}$ as a function of the magnetic field strength, $\mu_0 H$, at a constant microwave frequency 9.4 GHz, displayed at selected $\theta$ angles of the magnetic field. The vertical bars, marking the resonance field positions, are guides to the eyes for tracing the angular periodicity of the spectra. b) Dependence of the resonance field at room temperature on the orientation of the magnetic field, $H_{res}(\theta)$, upon its rotation in the plane spanned by the [001] and [110] axes. The angle $\theta$ is measured from the [001] axis as indicated in the inset, which depicts the single crystal under investigation prepared as a thin cylindrical disk. The solid black line shows a fit by uniaxial magnetocrystalline anisotropy [37–39], as described in the text.

**2-4. Magnetic textures observed by Lorentz transmission electron microscopy**

In order to observe magnetic structures in real space, LTEM measurements were carried out on (001) thin plates. We present the magnetic induction fields as deduced by transport-of-intensity equation (TIE) analyses for the compounds with Pd 7% (thickness $t \sim 130$ nm), Pd 4% ($t \sim 140$ nm), and Rh 8% ($t \sim 180$ nm) in **Figure 5**. The stripe pattern with a few hundred nm periods observed for all the compounds at room temperature and zero magnetic field (Figure 5a,d,g) corresponds to anisotropic helices with opposite helicities propagating along the [110] and [$\bar{1}$10] axes.

In the Pd 7% sample, square antiskyrmions appear uniformly over the plate under a magnetic field perpendicular to the specimen (Figure 5b). Note that high-density antiskyrmions are observed only after the sample plate is tilted under a magnetic field to initially create non-topological bubbles, and tilted back to the original perpendicular position as detailed in the caption for Figure 5.[14] Similar antiskyrmion lattices are observed at low temperatures down to 100 K (Figure 5c).

In the Pd 4% sample, square antiskyrmions and elliptically deformed skyrmions coexist under a magnetic field (Figure 5e). This coexistence state is also observed for $t \sim 170$ nm, 190 nm, and 220 nm, but a homogeneous antiskyrmion lattice is not formed. At the thickness of $t \sim 80$ nm, only elliptic skyrmions are observed. The (co)existence of skyrmions in spite of the anisotropic DMI indicates that the dipolar interaction is dominant over the DMI in this system.

In the Rh 8% sample, while dense elliptic skyrmions with mixed helicities are stabilized under a magnetic field (Figure 5h), no antiskyrmions are observed at any field for any sample thickness from 70 nm to 240 nm. As the temperature is lowered in zero field, the stripe pattern changes to large (~ several micrometers) domains with in-plane magnetizations (Figure 5j), which is consistent with the behavior of spin reorientation in the bulk magnetization measurement (Figure 3b). The in-plane domains are arranged in such a way that the magnetic flux is closed as indicated with white arrows. The transition from the anisotropic helices to the in-plane closure domains starts at higher temperatures for smaller thickness, e.g., 200 K for $t \sim$ 70 nm, 143 K for $t \sim$ 110 nm and 150 nm, and 130 K for $t \sim$ 180 nm. Furthermore, this transformation is accompanied by a coexistence region and a large thermal hysteresis; the stripe patterns remain partially at low temperatures, but once the change into the in-plane domains is complete, they hardly recover in a subsequent heating process (Figure S5).

In Figure 5f and 5i, enlarged views of an elliptic skyrmion are presented. The elliptic deformation is due to the cooperative interplay between the dipolar interaction and the anisotropic DMI inherent to the crystal structure with $D_{2d}$ or $S_4$ symmetry,[13,14,16] and hence distinct from elliptic skyrmions induced by artificial anisotropy.[40,41] From the observed ellipticity, the magnitude of DMI is roughly estimated to be $\sim$ 26% of the demagnetization energy.[13] In addition, the variation of the DMI with composition is negligible since the ellipticity of the skyrmion is comparable for the Pd 4% and Rh 8% compounds. Although the change in the DMI for these compounds investigated in this work is small, it might vary significantly in a wider compositional range, similarly to the Co/Pt bilayers where the interfacial DMI depends on the band filling of $5d$ orbitals.[42] To determine the anisotropic DMI more quantitatively and understand its compositional dependence comprehensively, further studies (e.g. Brillouin light scattering[43], spin-wave spectroscopy[44]) are necessary. Therefore, while the anisotropic DMI in this system is indeed important, we focus hereafter on the effects of larger and composition/temperature-dependent demagnetization energy and magnetic anisotropy on the formation of the topological spin textures.

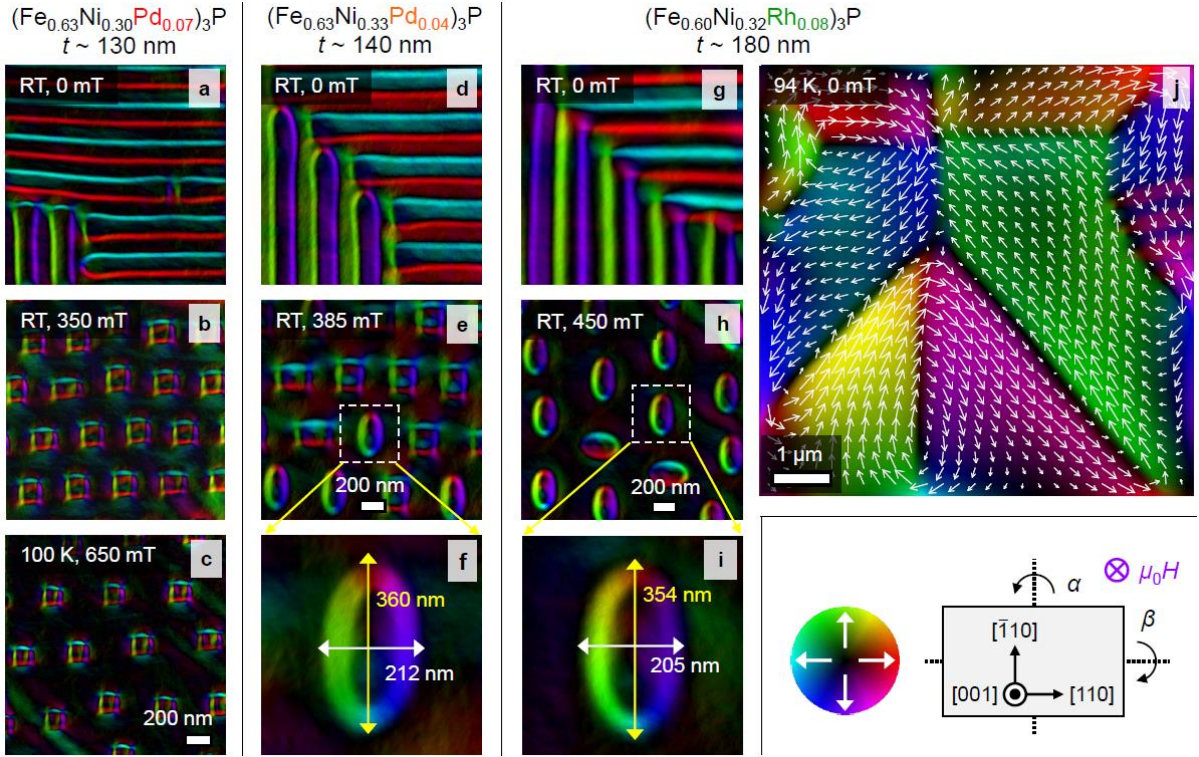

**Figure 5.** Magnetic structures observed by LTEM. a-j) Color mapping of in-plane magnetic induction fields deduced by transport-of-intensity equation (TIE) analyses of over- and under-focus LTEM images.[50] The color-coded wheel and the schematic figure of the experimental configuration (crystal axes of the thin plate and the external field) at the bottom right side are common for all the panels. The images of antiskyrmions and skyrmions (panels b,c,e,h) were obtained after tilting the plate under the external field and then back to the original position, as described in the following, where the tilt angles around the $[\bar{1}10]$ and $[110]$ axes are denoted as $α$ and $β$, respectively. a-c) Field images for $(Fe_{0.63}Ni_{0.30}Pd_{0.07})_3P$ with thickness $t \sim 130$ nm at (a) 295 K and 0 mT, (b) 295 K and 350 mT [process: 0 mT ($α \sim 0°$) → 350 mT ($α \sim 12°$) → 350 mT ($α \sim 0°$)], and (c) 100 K and 650 mT [process: ZFC → 100 K and 0 mT ($α \sim 0°$) → 550 mT ($α \sim 12°$) → 550 mT ($α \sim 0°$) → 650 mT ($α \sim 0°$)]. d-f) Field images for $(Fe_{0.63}Ni_{0.33}Pd_{0.04})_3P$ with $t \sim 140$ nm at (d) 295 K and 0 mT, and (e, f) 295 K and 385 mT [process: 0 mT ($α \sim 0°$, $β \sim 0°$) → 385 mT ($α \sim 2.4°$, $β \sim 1°$) → 385 mT ($α \sim 0°$, $β \sim 0°$)]. g-j) Field images for $(Fe_{0.60}Ni_{0.32}Rh_{0.08})_3P$ with $t \sim 180$ nm at (g) 295 K and 0 mT, (h, i) 295 K and 450 mT [process: 0 mT ($α \sim 0°$) → 450 mT ($α \sim 5°$) → 450 mT ($α \sim 0°$)], and (j) 94 K and 0 mT. Panels f and i display enlarged views of the elliptic skyrmions shown in panels e and h, respectively. The ellipticity (ratio of the length of the major axis to that of the minor axis) of the skyrmion in panels f and i are 1.70 and 1.73, respectively. The magnetic induction fields in the in-plane domains (panel j) are also indicated with white arrows.

## 2-5. Stability of antiskyrmions governed by demagnetization energy and uniaxial anisotropy

To understand the effect of the thickness-dependent dipolar interactions quantitatively, we estimate demagnetization energy $E_d$ for the thin plates using the following theoretical equation for a helical stripe (Bloch wall type) with a period of $\lambda$ and a film thickness of $t$,[19,45]

$$E_d = \frac{\mu_0 M_s^2}{2} f\left(\frac{2\pi t}{\lambda}\right) \qquad (2)$$

where the function $f(x)$ is defined as $f(x) = (1 - e^{-x})/x$. For small $x$ (i.e. $t \ll \lambda$), $f(x)$ is simplified as $f(x) \approx 1 - x/2$, while $f(x) \approx 1/x$ for large $x$ (i.e. $t \gg \lambda$). **Figure 6**a shows $\lambda(t)$ of the anisotropic helices at room temperature and zero field for the composition with Pd 7%, Pd 4%, and Rh 8%. The thickness dependence is well described by the Kittel's law ($\lambda \propto t^{1/2}$),[46] which is attributed to the competition between demagnetization energy and domain wall energy. The calculated $E_d$ by using Equation (2) at room temperature for each compound is plotted against $t$ in Figure 6b, which increases monotonically with decreasing thickness.

Finally, we map out the magnetic objects observed by LTEM onto the plane of $K_u$ and $E_d$ in Figure 6c. Here, anisotropic helices and in-plane domains at zero field, and antiskyrmions/skyrmions under magnetic fields are plotted all together. The $K_u$–$E_d$ phase diagram indicates that antiskyrmions are stabilized when $K_u$ is positive, sufficiently large and comparable to $E_d$. On the other hand, when $K_u$ is very small as compared to $E_d$, as in the cases of Rh doping, or Pd doping with small thickness, elliptic skyrmions become more dominant. This result proves that the strong dipolar interaction destabilizes the antiskyrmions with magnetic volume charge, and favors the Bloch skyrmions without it. Therefore, uniaxial anisotropy and dipolar interaction are both important ingredients for stabilizing antiskyrmions in $S_4$ magnets with anisotropic DMI (and probably $D_{2d}$ magnets as well). The significant role of uniaxial anisotropy and dipolar interaction for stabilizing antiskyrmions has been also demonstrated in multilayers.[15]

The phase diagram in Figure 6c also describes the temperature-induced change from the anisotropic helices to the in-plane domains when the sign of $K_u$ changes to negative in the case of the Rh 8% doping (although initially formed anisotropic helices at room temperature with $K_u > 0$ can survive to the region of $K_u < 0$ as a metastable state). The phase boundary between the helices and the in-plane domains shows a positive slope, indicating that the dipolar interaction also contributes to the spin reorientation. The easy-plane anisotropy may also give rise to more interesting in-plane topological magnetism such as bimeron.[47,48]

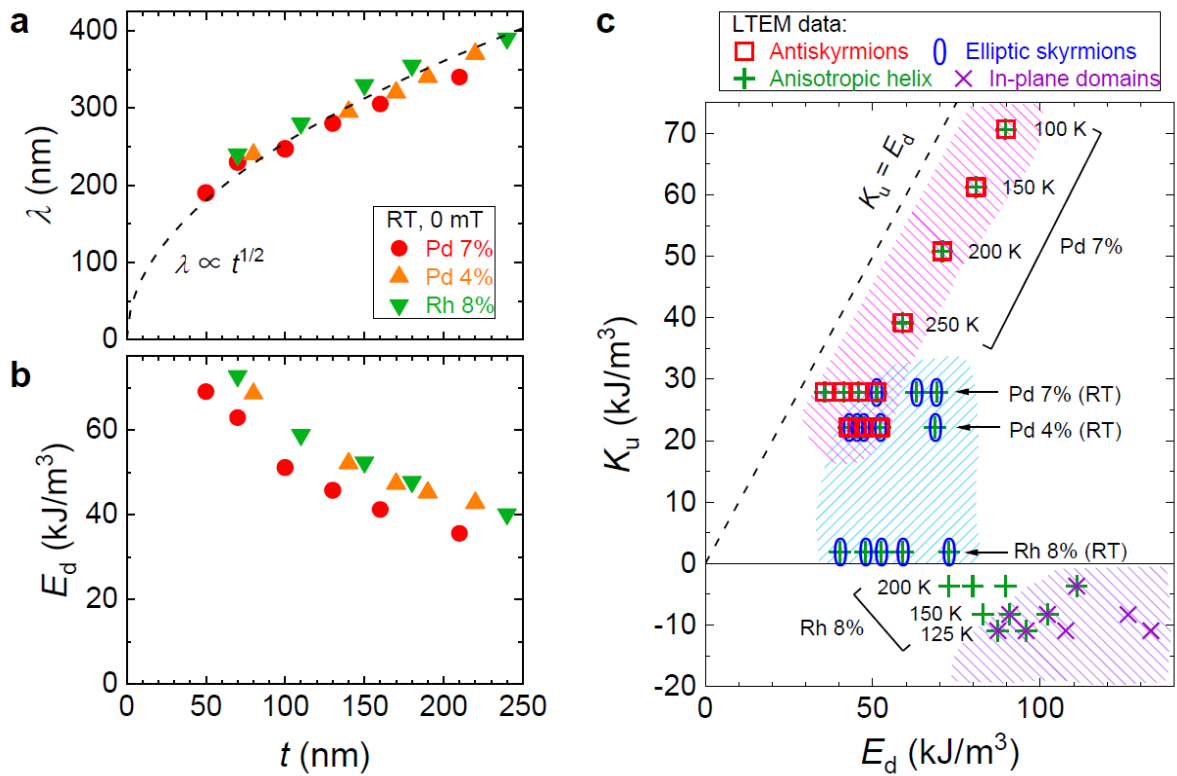

**Figure 6.** Demagnetization energy and magnetic phase diagram. a, b) Thickness ($t$) dependence of (a) magnetic periodicity $\lambda$ of the anisotropic helix and (b) demagnetization energy $E_d$ at room temperature for $(Fe_{0.63}Ni_{0.30}Pd_{0.07})_3P$, $(Fe_{0.63}Ni_{0.33}Pd_{0.04})_3P$, and $(Fe_{0.60}Ni_{0.32}Rh_{0.08})_3P$. c) Magnetic phase diagram observed by LTEM on the plane of uniaxial anisotropy constant $K_u$ and $E_d$. All the spin textures are plotted together, which include anisotropic helix and in-plane domains at zero field, and antiskyrmions and skyrmions under fields, for the Pd 7%, Pd 4%, and Rh 8% compounds with various thicknesses shown in panel b. Low-temperature data for Pd 7% ($t = 130$ nm) and Rh 8% ($t \leq 180$ nm) are also included. The data points for Pd 7% are reproduced in part from our earlier work.[14] (Copyright 2021, Springer Nature).

## 2-6. Micromagnetic simulations

The experimental observation of the antiskyrmions stabilized by uniaxial anisotropy is in qualitative agreement with our micromagnetic simulations, presented in **Figure 7**. In Figure 7a, we show simulation results of magnetic textures at various $K_u$ and external magnetic fields with a constant dipolar energy density of $E_d = 66$ kJ/m$^3$ (note that the horizontal axis is different from that in Figure 6c). We start from an initially triangular antiskyrmion lattice and relax it at finite temperature. As a result, we obtain a field-polarized ferromagnetic state at large magnetic fields $\mu_0 H \geq 500$ mT, in-plane states at low anisotropy $K_u \leq 30$ kJ/m$^3$, and antiskyrmions at large anisotropies $K_u \geq 70$ kJ/m$^3$ $\approx E_d$ of the order of the dipolar energy density $E_d$ and average external magnetic fields $\mu_0 H \leq 375$ mT. This antiskyrmion phase pocket is surrounded by a belt of dipolar stabilized skyrmions, including a crossover regime where skyrmions and antiskyrmions coexist. A particularly instructive example is displayed in the panel for $\mu_0 H = 300$ mT and $K_u = 80$ kJ/m$^3$ (enlarged view is shown in Figure 7c), which contains the characteristically distorted anti-/skyrmions: two square-shaped antiskyrmions in the lower row and in the upper row two elliptical skyrmions with opposite handedness and consequently, perpendicular elongation axes. Moreover, we checked that for thicker films the onset of the antiskyrmion phase shifts to lower anisotropies, in agreement with both our experimental observations and the argument that DMI-stabilized antiskyrmions benefit from the larger volume fraction.

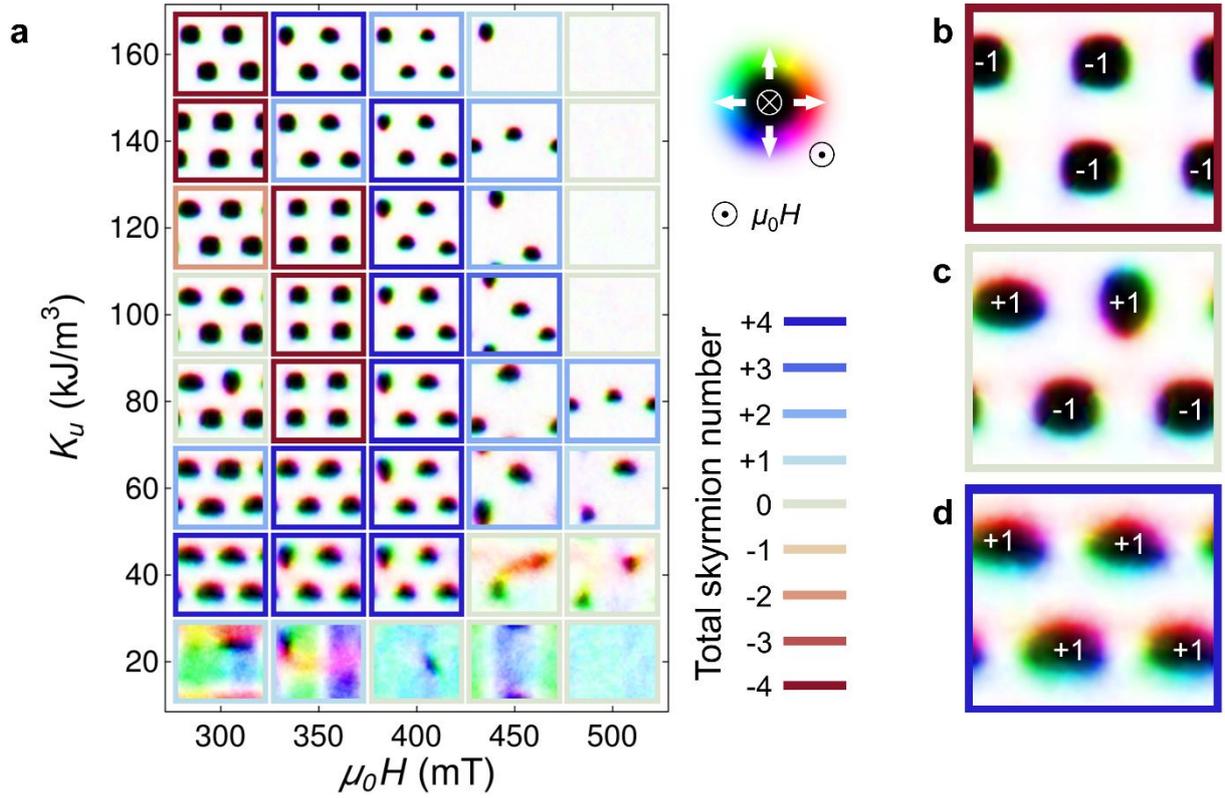

**Figure 7.** Thickness-averaged magnetization as obtained from three-dimensional micromagnetic simulations for a 100 nm thick film. a) Simulated magnetic textures on the plane of uniaxial anisotropy energy $K_u$ and external magnetic field $\mu_0 H$. The color encodes the orientation of the magnetization with white (black) indicating the $+z$ ($-z$) component. The color of the frame of every panel encodes the total skyrmion number from +4 (blue, four skyrmions) to -4 (red, four antiskyrmions). The dipolar energy density in the simulations is $E_d = 66$ kJ/m$^3$. b-c) Enlarged views of the magnetic textures at $\mu_0 H = 300$ mT and (b) $K_u = 140$ kJ/m$^3$, (c) 80 kJ/m$^3$, (d) 40 kJ/m$^3$, where the skyrmion number for each magnetic object is also indicated (+1 for skyrmion, -1 for antiskyrmion).

## 3. Conclusion

We have systematically studied the magnetic properties of schreibersite (Fe,Ni)$_3$P with $S_4$ symmetry by performing magnetometry, FMR spectroscopy, LTEM, and micromagnetic simulations, demonstrating that the magnetic anisotropy can be finely controlled by the composition to stabilize antiskyrmions. The strong easy-plane anisotropy of Fe$_3$P is rapidly suppressed by partial Ni substitution, and when additionally doped with a small amount of Pd, the magnetic anisotropy switches to an easy-axis type, leading to the formation of antiskyrmions. The Rh-doped compound exhibits a temperature-induced spin reorientation transition, which is directly observed by LTEM. The phase diagram regarding the magnetic textures clearly shows stable regions of antiskyrmions and elliptic skyrmions on the plane of the uniaxial anisotropy and the demagnetization energy. These findings unveil that easy-axis type uniaxial magnetic anisotropy and dipolar interaction with appropriate balance are both necessary to stabilize antiskyrmions, and will help to design new antiskyrmion systems towards applications in spintronics.

## 4. Methods

*Sample preparation*: Bulk single crystals of $M_3P$ (M: Fe, Ni, Ru, Rh, and Pd) were synthesized by a self-flux method from the initial molar ratio of M : P ~ 3.5 : 1 (Note S1 and Table S1). Chemical compositions of the obtained crystals were determined by the energy dispersive X-ray analysis (Table S1). Lattice constants of $M_3P$ with the non-centrosymmetric tetragonal structure were determined by powder X-ray diffraction and Rietveld analysis (Note S2, Figure S1, and Figure S2).

*Magnetization measurement*: For magnetization measurements, single crystals were cut into a 1-mm-scale rectangle with flat (110) and (001) surfaces and approximately the same length along the [110] and [001] axes (Figure S3). Magnetization measurements were carried out using a superconducting quantum interference device magnetometer (MPMS3, Quantum Design) equipped with an oven option.

*FMR measurement*: Angular dependent ferromagnetic resonance (FMR) experiments were performed for the temperature range 250 K ≤ $T$ ≤ 380 K, using a Bruker ELEXSYS E500 CW X-band (9.4 GHz) spectrometer. The sample of $(Fe_{0.63}Ni_{0.30}Pd_{0.07})_3P$ was in a continuous nitrogen gas-flow cryostat with temperature stability of about 1 K. The measurements were performed on a cylindrical thin disk (diameter/thickness = 5), whose plane contains both the tetragonal [001] axis and the [110] axis perpendicular to it. The orientation of the sample was controlled by a programmable goniometer in 5° steps during a full rotation of the magnetic field in the plane of the cylindrical disk at selected temperatures: 250 K, 300 K, 350 K and 380 K.

To obtain the uniaxial anisotropy parameter $K_u$ at these temperatures, the angular dependence of the resonance field was fitted using the Smit-Suhl formula,[49] similarly to the procedure followed in Ref.[38].

*LTEM measurement*: For Lorentz transmission electron microscopy (LTEM) measurements, (001) thin plates with various thickness were thinned from bulk single crystals by a focused ion beam (FIB) system (NB5000, Hitachi); $t \sim$ 70, 110, 150, 180, and 240 nm for $(Fe_{0.60}Ni_{0.32}Rh_{0.08})_3P$; $t \sim$ 80, 140, 170, 190, and 220 nm for $(Fe_{0.63}Ni_{0.33}Pd_{0.04})_3P$; $t \sim$ 50, 70, 100, 130, 160, and 210 nm for $(Fe_{0.63}Ni_{0.30}Pd_{0.07})_3P$. LTEM measurements were performed with a transmission electron microscope (JEM-2100F, JEOL) equipped with a double-tilt liquid-nitrogen holder (Gatan 636) and a double-tilt heating holder (Protochips: Fusion select). External magnetic fields applied to the (001) plates were obtained by tuning the objective lens current of JEM-2100F, which are parallel to the incident electron beam. The distribution of in-plane magnetic induction fields was obtained by a transport of intensity equation (TIE) analysis using over- and under-focus LTEM images.[50]

*Micromagnetic simulations:* For the theoretical description of the magnetization we consider the same continuum model as in Ref.[14], i.e., including the magnetic stiffness $A_{ex}$, $S_4$-symmetric Dzyaloshinskii-Moriya interaction $D$ (DMI), uniaxial anisotropy $K_u$, Zeeman field $H$, and dipolar interactions due to the saturation magnetization $M_s$. For the numerical implementation, we use our modified version of MuMax3[51,52] in which we flipped the sign of the derivatives in the DMI-field in one spatial direction. Thus, the DMI favors right-handed helices in the horizontal direction and left-handed helices in the vertical direction. We choose the micromagnetic parameters as $A_{ex} = 8.1$ pJ/m, $D = 0.2$ mJ/m$^2$, and $M_s = 600$ kA/m. For obtaining the results in Figure 7, we first relax a triangular lattice of antiskyrmions at zero temperature at some convenient values of the magnetic field and anisotropy. The simulated system with periodic boundary conditions in the x-y-plane measures 800 nm x 700 nm x 100 nm, discretized on 128 x 112 x 32 lattice sites, and contains 4 antiskyrmions. In a second step, we use this texture as the starting point for simulations at finite temperature $T$ ($T = 600$ K) for a timespan $t$ after which the textures appear to be in quasi-equilibrium ($t = 10$ ns). This method yields an approximation for the energetically most favorable texture which otherwise is a complex high-dimensional optimization problem with many local minima. The resulting thickness-averaged magnetization is shown in Figure 7 together with the skyrmion number which is the two-dimensional winding number[53] up to a sign. The dipolar energy density $E_d = 66$ kJ/m$^3$ is determined by deterministically finding the optimized helical wavelength at every value of the anisotropy which turns out to be qualitatively constant in the regime of interest.

**Acknowledgements**

The authors thank R. Arita for fruitful discussions. We are grateful to V. Tsurkan and L. Prodan for preparation of the cylindrical thin disk for FMR measurements. This work was supported by JSPS Grant-in-Aids for Scientific Research (Grant No. 19H00660, 20K15164) and JST CREST (Grant No. JPMJCR20T1 and JPMJCR1874) and via the DFG Priority Program SPP2137, Skyrmionics, under Grant Nos. KE 2370/1-1. M.H. and H.-A.K.v.N. acknowledge funding within the joint RFBR-DFG research project Contracts No. 19-51-45001 and No. KR2254/3-1. J.M. was supported as Humboldt/JSPS International Research Fellow (19F19815) and by the Alexander von Humboldt Foundation as a Feodor Lynen Return Fellow.

# Supporting Information

Doping control of magnetic anisotropy for stable antiskyrmion formation in schreibersite (Fe,Ni)$_3$P with $S_4$ symmetry

*Kosuke Karube\*, Licong Peng, Jan Masell, Mamoun Hemmida, Hans-Albrecht Krug von Nidda, István Kézsmárki, Xiuzhen Yu, Yoshinori Tokura, Yasujiro Taguchi*

**Note S1. Sample preparations**

Bulk single crystals were prepared by the following self-flux method. Pure metal M (Fe, Ni, Ru, Rh, Pd) powders (except for wire cuts of Pd) and red phosphorous P were loaded into a boron nitride (BN) crucible in a molar ratio of M : P ~ 3.5 : 1, with a total mass of ~ 3 g. Details of the loaded compositions are given in Table S1. The excess amount of M acts as a flux for the growth of M$_3$P and avoids formation of M$_2$P. The BN crucible was sealed in an evacuated quartz tube and heated slowly in a standard box furnace to 1100 ~ 1200°C over 3 ~ 4 days, held for 12 ~ 24 hours, and then cooled to room temperature over 12 ~ 24 hours. At this preliminary reaction stage, the ingot contains polycrystalline M$_3$P and other phases (M-rich cubic phase and M$_2$P). The ingot was then crushed into small pieces, sealed together with the BN crucible again in an evacuated quartz tube, heated above the melting point and then slowly cooled in a standard box furnace or a Bridgman furnace (see Table S1 for details of temperatures and cooling rates). After this slow cooling, single crystals of M$_3$P with metallic luster ranging in size from ~ 0.5 mm to ~ 2 mm were found in the ingot. It was found that larger single crystals could be obtained by Bridgman growth. Since the flux (mixture of M-rich cubic phase and M$_3$P) remained in the ingot and could not be completely removed by centrifugation, the ingot was finally crushed and cut to separate the single crystals of M$_3$P.

The final chemical composition of the obtained M$_3$P crystals was determined by an energy dispersive X-ray (EDX) spectrometer (Bruker, XFlash6) equipped with a scanning electron microscope (JEOL, JSM-6701F). The standardless $\varphi(\rho z)$ method was used to quantify the EDX spectra. The results of the EDX analysis are summarized in Table S1.

**Table S1.** Summary of single crystal growth methods and EDX results of $M_3P$

| Composition | Loading mol ratio | Single crystal growth process after the second melting | EDX results |
|---|---|---|---|
| $Fe_3P$ | Fe : P = 3.6 : 1 | Slow cool from 1160°C to 1050°C at a rate of 0.4°C/h in a box furnace | $Fe_{3.026(3)}P_{0.974(3)}$ |
| $(Fe_{0.82}Ni_{0.18})_3P$ | Fe : Ni : P = 2.8 : 0.7 : 1 | Slow transfer from the hot zone (1030°C) to the cold zone (950°C) in a Bridgman furnace at a rate of 0.5 mm/h for 11 days | $Fe_{2.489(8)}Ni_{0.534(5)}P_{0.977(6)}$ |
| $(Fe_{0.63}Ni_{0.37})_3P$ | Fe : Ni : P = 2.05 : 1.45 : 1 | Slow transfer from the hot zone (1025°C) to the cold zone (950°C) in a Bridgman furnace at a rate of 0.4 mm/h for 13 days | $Fe_{1.911(4)}Ni_{1.103(7)}P_{0.985(3)}$ |
| $(Fe_{0.53}Ni_{0.47})_3P$ | Fe : Ni : P = 1.75 : 1.75 : 1 | Slow cool from 1020°C to 960°C at a rate of 0.2°C/h in a box furnace | $Fe_{1.609(7)}Ni_{1.409(5)}P_{0.982(2)}$ |
| $(Fe_{0.34}Ni_{0.66})_3P$ | Fe : Ni : P = 1.15 : 2.35 : 1 | Slow cool from 990°C to 950°C at a rate of 0.2°C/h in a box furnace | $Fe_{1.025(4)}Ni_{1.987(8)}P_{0.988(5)}$ |
| $(Fe_{0.59}Ni_{0.32}Ru_{0.09})_3P$ | Fe : Ni : Ru : P = 2.0 : 1.2 : 0.3 : 1 | Slow transfer from the hot zone (990°C) to the cold zone (930°C) in a Bridgman furnace at a rate of 0.5 mm/h for 10 days | $Fe_{1.776(7)}Ni_{0.967(8)}Ru_{0.267(2)}P_{0.989(2)}$ |
| $(Fe_{0.60}Ni_{0.32}Rh_{0.08})_3P$ | Fe : Ni : Rh : P = 2.0 : 1.2 : 0.3 : 1 | Slow transfer from the hot zone (990°C) to the cold zone (930°C) in a Bridgman furnace at a rate of 0.5 mm/h for 10 days | $Fe_{1.795(3)}Ni_{0.975(4)}Rh_{0.239(1)}P_{0.991(4)}$ |
| $(Fe_{0.63}Ni_{0.33}Pd_{0.04})_3P$ | Fe : Ni : Pd : P = 2.0 : 1.2 : 0.3 : 1 | Slow transfer from the hot zone (950°C) to the cold zone (900°C) in a Bridgman furnace at a rate of 0.5 mm/h for 6 days | $Fe_{1.911(4)}Ni_{0.997(4)}Pd_{0.121(1)}P_{0.971(3)}$ |
| $(Fe_{0.63}Ni_{0.30}Pd_{0.07})_3P$ | Fe : Ni : Pd : P = 1.8 : 1.1 : 0.6 : 1 | Slow transfer from the hot zone (970°C) to the cold zone (910°C) in a Bridgman furnace at a rate of 0.5 mm/h for 9 days | $Fe_{1.897(7)}Ni_{0.892(8)}Pd_{0.217(3)}P_{0.994(5)}$ |
| $(Fe_{0.62}Ni_{0.29}Pd_{0.09})_3P$ | Fe : Ni : Pd : P = 1.75 : 1.05 : 0.7 : 1 | Slow transfer from the hot zone (970°C) to the cold zone (910°C) in a Bridgman furnace at a rate of 0.5 mm/h for 9 days | $Fe_{1.877(6)}Ni_{0.898(4)}Pd_{0.259(2)}P_{0.966(3)}$ |

**Note S2. Structural characterizations**

The non-centrosymmetric tetragonal crystal structure of M$_3$P with the space group of $I\bar{4}$ (No. 82, $S_4^2$) was confirmed by powder X-ray diffraction (Rigaku, RINT-TTR III) with Cu K$\alpha$ radiation as shown in Figures S1 and S2. The lattice constants $a$ and $c$ were determined by Rietveld analysis using the RIETAN-FP program, [S1] and are summarized in Table 1 in the main text. In the M$_3$P structure, there are three different crystallographic M sites (M1, M2, M3). It is difficult to determine the exact site occupancy in the three sites by X-ray diffraction because of the similar scattering powers of Fe and Ni, and the small concentration of 4$d$ metals. In the Rietveld calculations shown in Figures S1 and S2, we assume that M is evenly distributed over the three sites, and good fits were obtained. Previous synchrotron radiation diffraction study on (Fe,Ni)$_3$P reported that Fe and Ni preferentially occupy the M1 and M3 sites, respectively, while the M2 site is randomly occupied by both Fe and Ni.[S2] However, the refinements after setting this preferential site occupancy for Fe and Ni did not show further improvement.

The (110) and (001) planes of the single crystals were determined by X-ray back-reflection Laue photography (Rigaku, RASCO-BL II). In order to confirm the quality of the single crystals, X-ray diffraction on the (001) surface of a single crystal of (Fe$_{0.63}$Ni$_{0.30}$Pd$_{0.07}$)$_3$P was carried out. As shown in Figure S3, the full width at half maximum (FWHM) of the rocking ($\theta$) scan of the (004) peak was ~ 0.08°. The single peak and the narrow FWHM indicate that this single crystal is of high quality.

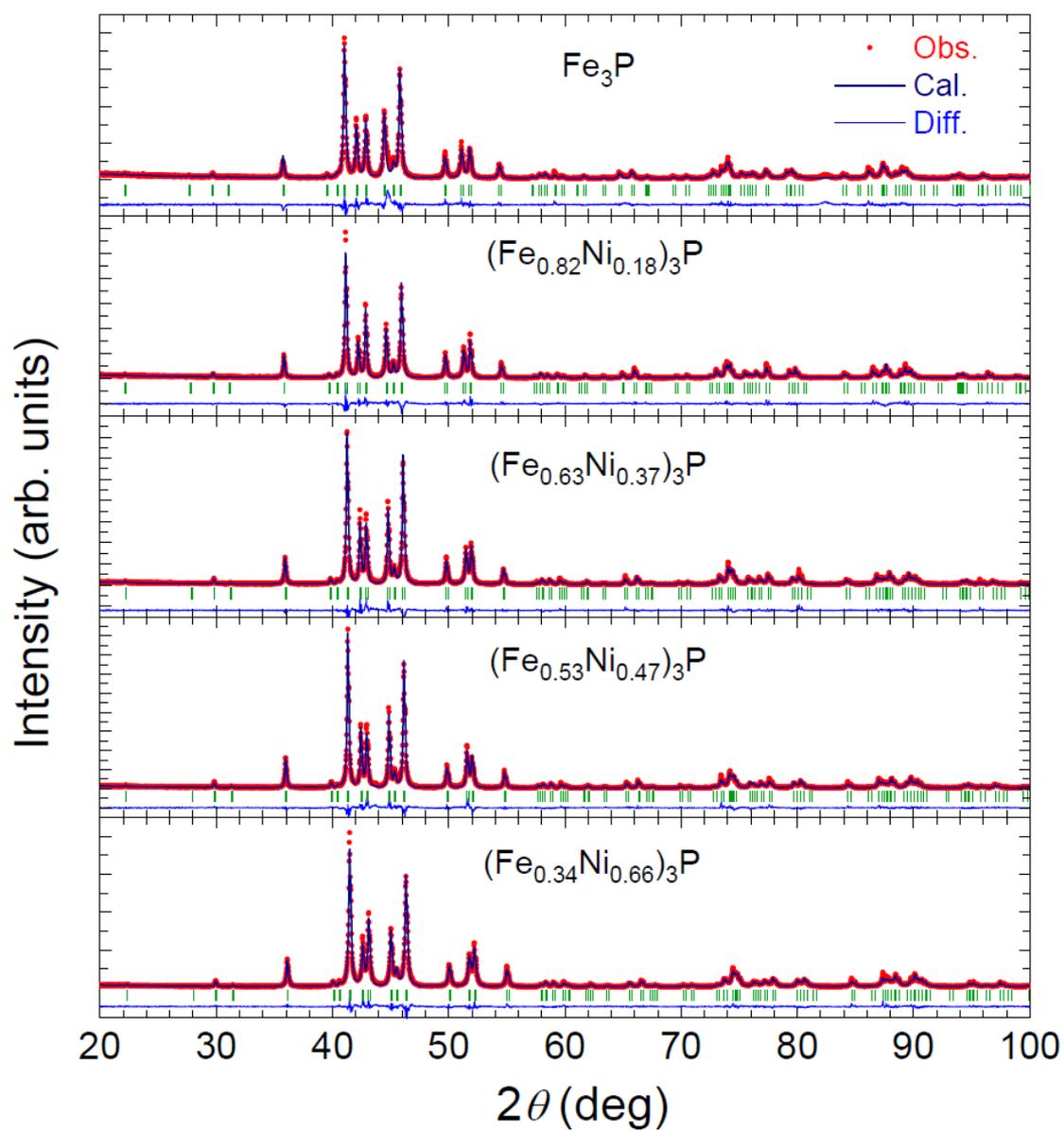

**Figure S1.** Powder X-ray diffraction patterns and Rietveld analysis of $(Fe_{1-x}Ni_x)_3P$ at room temperature.

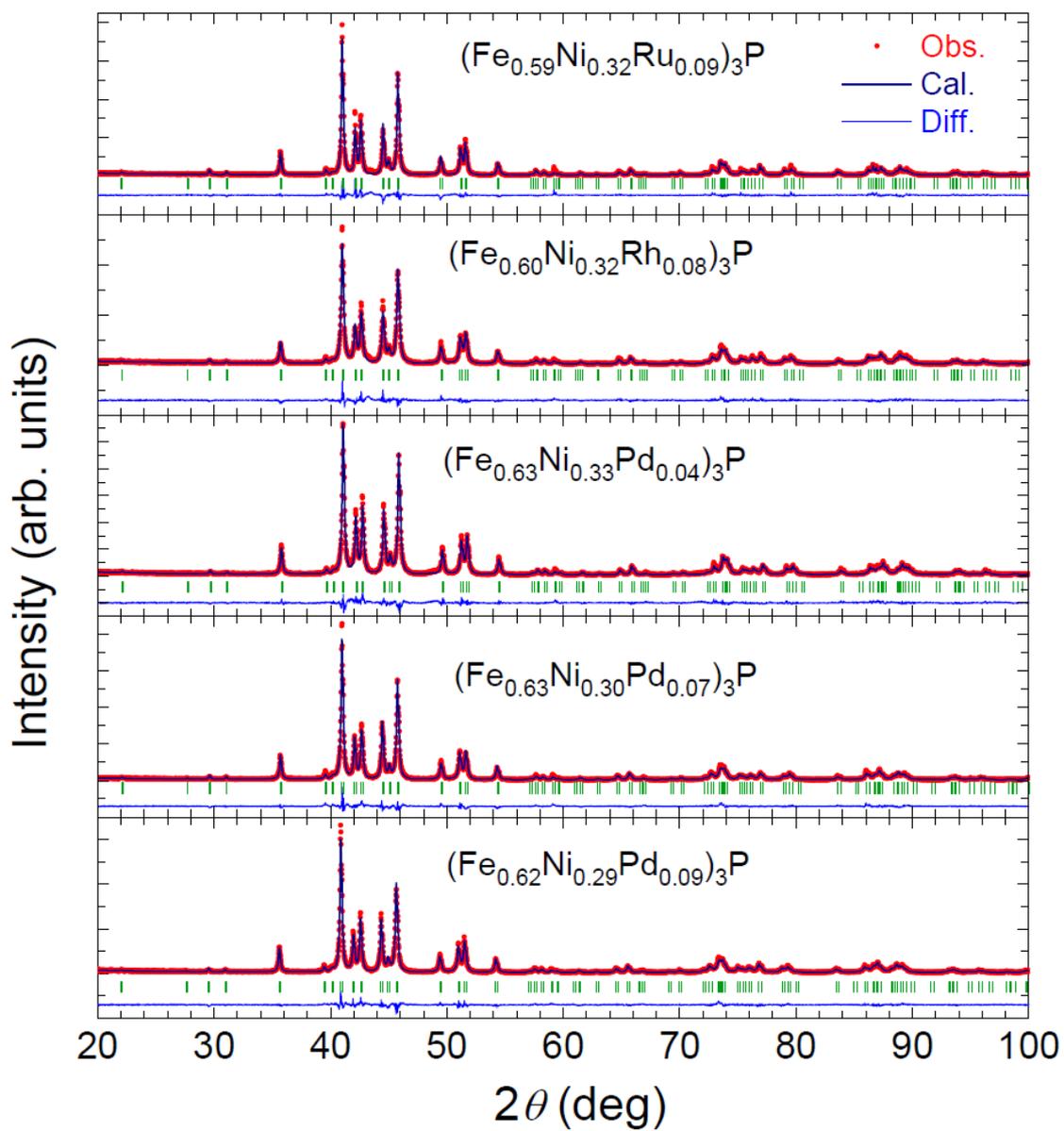

**Figure S2.** Powder X-ray diffraction patterns and Rietveld analysis of compounds doped with 4*d* transition metals at room temperature.

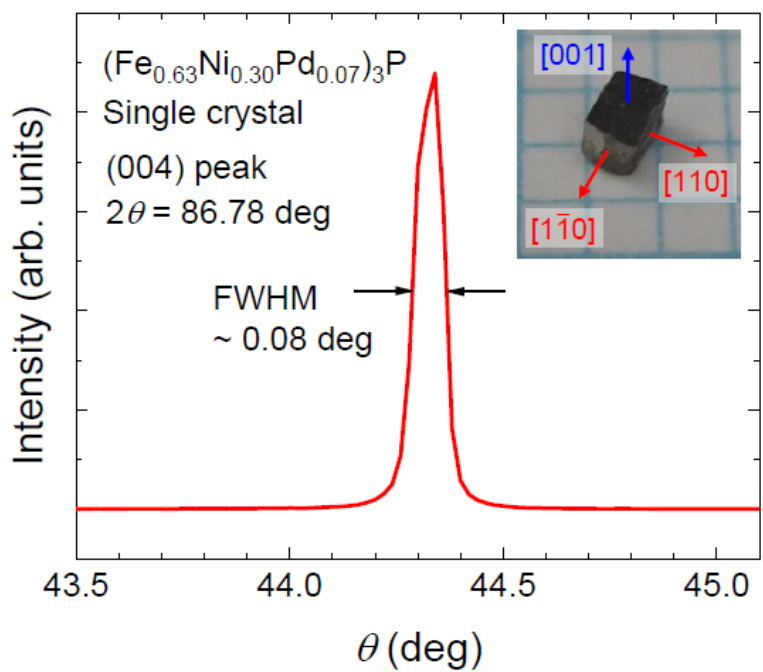

**Figure S3.** Rocking curve of the (004) X-ray diffraction peak (at $2\theta = 86.78$ deg) at room temperature on the surface of the single crystal of $(Fe_{0.63}Ni_{0.30}Pd_{0.07})_3P$. The inset shows the photograph of the single crystal on a mm-scale grid sheet.

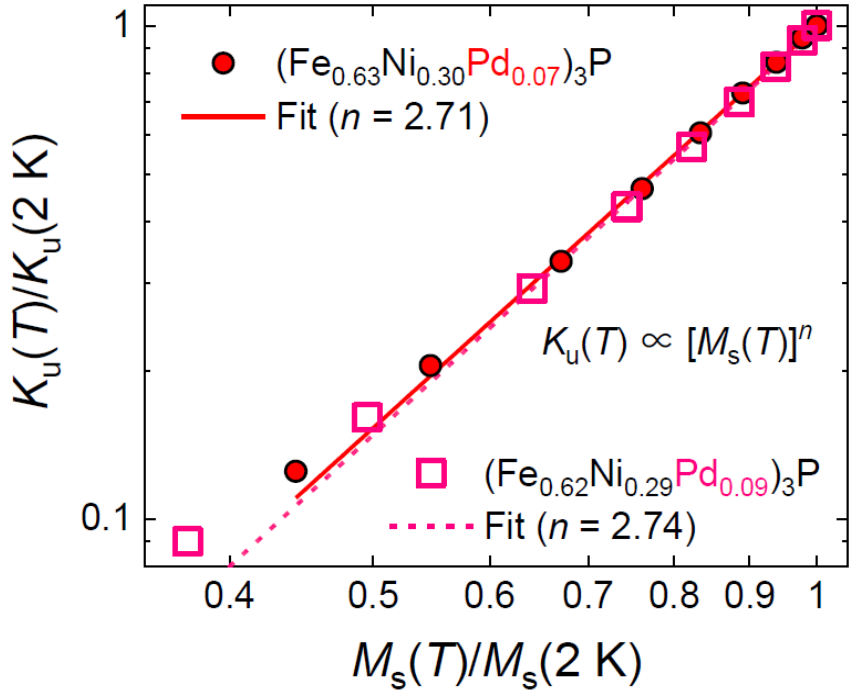

**Figure S4.** The relation between normalized uniaxial anisotropy constant $K_u(T)/K_u(2\ K)$ and the saturation magnetization $M_s(T)/M_s(2\ K)$ for $(Fe_{0.63}Ni_{0.30}Pd_{0.07})_3P$ and $(Fe_{0.62}Ni_{0.29}Pd_{0.09})_3P$ at various temperatures is plotted on a logarithmic scale. Power-law fits are indicated by lines.

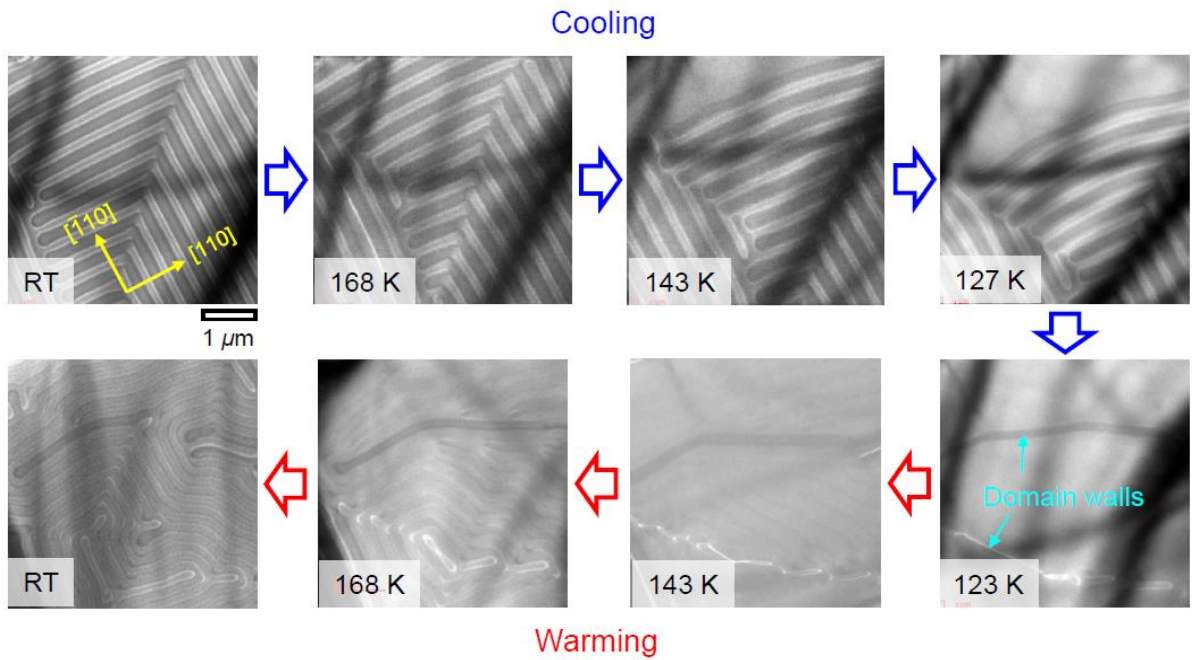

**Figure S5.** Temperature variations of LTEM images for $(Fe_{0.60}Ni_{0.32}Rh_{0.08})_3P$ with a thickness of $t \sim 110$ nm at zero field. The cooling process from room temperature (RT) down to 123 K and the subsequent warming process back to RT are presented. The domain walls of the in-plane ferromagnetic state at 123 K are indicated with the light blue arrows.